%% file: main.tex
\documentclass[sigplan,10pt,screen]{acmart}

\renewcommand\footnotetextcopyrightpermission[1]{}

\input{utils}
\begin{document}

% \title{The Name of the Title Is Hope}
\title{\sys: Transparent I/O Offloading for High-Density Serverless Computing}
% \subtitle{Separation of Powers: Transparent Compute-IO Separation in Serverless Clouds}

% %%
% % The "author" command and its associated commands are used to define
% % the authors and their affiliations.
% % Of note is the shared affiliation of the first two authors, and the
% % "authornote" and "authornotemark" commands
% % used to denote shared contribution to the research.
% \affiliation{%
\author{JooYoung Park}
\email{jooyoung001@e.ntu.edu.sg}
\affiliation{%
  \institution{NTU Singapore}
  \country{Singapore}
}

\author{Kevin Nguetchouang}
\email{kevin.nn@ntu.edu.sg}
\affiliation{%
  \institution{NTU Singapore}
  \country{Singapore}
}

\author{Jovan Stojkovic}
\email{jovan.stojkovic@utexas.edu}
\affiliation{%
  \institution{UT Austin}
  \country{USA}
}

\author{Likun Zhang}
\email{likun001@e.ntu.edu.sg}
\affiliation{%
  \institution{NTU Singapore}
  \country{Singapore}
}

\author{Riccardo Mancini}
\email{mancio@amazon.co.uk}
\affiliation{%
  \institution{AWS}
  \country{United Kingdom}
}

\author{Marco Cali}
\email{calmar91ct@gmail.com}
\affiliation{%
  \institution{AWS}
  \country{United Kingdom}
}

\author{Dmitrii Ustiugov}
\email{dmitrii.ustiugov@ntu.edu.sg}
\affiliation{%
  \institution{NTU Singapore}
  \country{Singapore}
}

\renewcommand{\shortauthors}{Park et al.}

\settopmatter{printfolios=true,printacmref=false}
\pagestyle{plain}

\input{sec/0_abstract}

\maketitle

\renewcommand{\emph}[1]{\textit{#1}}

\input{sec/1_introduction}
\input{sec/2_background}
\input{sec/3_characterization}

\input{sec/4_design}

\input{sec/6_evaluation}
\input{sec/5_related}
\input{sec/7_conclusion}

\begin{acks}

The authors thank the members of the HyScale lab at NTU Singapore for their constructive discussions and feedback on this work. This project is supported by the Ministry of Education, Singapore, under its Academic Research Funds Tier 2 MOE-T2EP20124-0002. 
\end{acks}

\bibliographystyle{ACM-Reference-Format}
\bibliography{./bibcloud/gen-abbrev,misc,dblp}

\end{document}

%% file: utils.tex
\usepackage{glossaries}
\makeglossaries
\usepackage{tikz}

\usepackage{mathtools}

\usepackage{subcaption}
\usepackage{xspace}
\usepackage{ulem}
\usepackage{xcolor}

\usepackage{listings}

\newcommand{\sys}{Nexus\xspace}

\newcommand*{\RELEASE}{} 
\ifdefined\RELEASE
  \newcommand{\ignore}[1]{}
  \newcommand{\fixme}[1]{}
  \newcommand{\dmi}[1]{}
  \newcommand{\cz}[1]{}
  \newcommand{\jy}[1]{}
  \newcommand{\kev}[1]{}
  \newcommand{\jovan}[1]{}
  \newcommand{\TODO}[1]{}
  \newcommand{\todo}[1]{}

\else
  \newcommand{\ignore}[1]{}
  \newcommand{\fixme}[1]{{\textcolor{red}{[~FIXME:~#1~]}}}
  \newcommand{\dmi}[1]{{\textcolor{blue}{[~D:~#1~]}}}
  \newcommand{\cz}[1]{{\textcolor{violet}{[~CZ:~#1~]}}}
  \newcommand{\jy}[1]{{\textcolor{teal}{[~JY:~#1~]}}}
  
  \newcommand{\kev}[1]{{\textcolor{cyan}{[~K:~#1~]}}}
  \newcommand{\jovan}[1]{{\textcolor{gray}{[~JO:~#1~]}}}
  \newcommand{\TODO}[1]{{\textcolor{red}{TODO:~#1}}}
  \newcommand{\todo}[1]{{\textcolor{red}{TODO:~#1}}}

\fi

\definecolor{codegreen}{rgb}{0,0.6,0}
\definecolor{codegray}{rgb}{0.5,0.5,0.5}
\definecolor{codepurple}{rgb}{0.58,0,0.82}
\definecolor{backcolour}{rgb}{1,1,1}

\lstdefinestyle{mystyle}{
    backgroundcolor=\color{backcolour},   
    commentstyle=\color{codegreen},
    keywordstyle=\color{magenta},
    numberstyle=\tiny\color{codegray},
    stringstyle=\color{codepurple},
    basicstyle=\ttfamily\footnotesize,
    breakatwhitespace=false,         
    breaklines=true,                 
    captionpos=b,                    
    keepspaces=true,                 
    numbers=left,                    
    numbersep=5pt,                  
    showspaces=false,                
    showstringspaces=false,
    showtabs=false,                  
    tabsize=2
}
\newcommand{\circled}[1]{\raisebox{.5pt}{\textcircled{\raisebox{-.9pt} {#1}}}}

%% file: sec/0_abstract.tex
\begin{abstract}

Serverless platforms rely on KVM-based virtual machines (VMs) to ensure strong isolation and compatibility with the rich ecosystem of libraries and images. However, current architectures tightly couple application logic with I/O processing, forcing every VM to duplicate a heavyweight communication fabric—comprising cloud SDKs, RPC frameworks, and the TCP/IP stack. 
% \sout{Our analysis reveals this duplication consumes over 25\% of a function's memory footprint and 60\% of user-space CPU cycles, severely limiting deployment density.}
Our analysis reveals this duplication consumes over 25\% of a function's memory footprint, and may double the CPU cycles in VMs compared to bare-metal execution.
Prior attempts to mitigate this using WebAssembly or library OSes sacrifice compatibility, forcing developers to migrate code and dependencies to low-level languages.

We introduce \sys, a serverless-native KVM hypervisor that transparently decouples compute from I/O. \sys intercepts the communication fabric at the high-level API boundary, remoting it to a shared host backend via zero-copy shared memory. This completely extracts the infrastructure tax from the guest without requiring any user code modifications. Furthermore, this structural separation unlocks asynchronous optimizations: by leveraging ingress routing hints, \sys completely overlaps input payload prefetching with VM restoration and safely defers output writes off the critical path. Compared to the AWS Firecracker baseline, \sys reduces node-level CPU and memory consumption by up to 44\% and 31\%, respectively, and increases deployment density by 18\% atop TCP and 37\% atop RDMA, demonstrating that KVM-based serverless architectures can achieve high density while retaining ecosystem compatibility.

\end{abstract}

%% file: sec/1_introduction.tex
\section{Introduction}

In serverless clouds, application developers offload deployment and data management to the provider, focusing only on their application logic, which is defined as a set of functions, instances of which the provider scales on demand. However, this model is economically viable only if the providers can maximize deployment density by colocating hundreds to thousands of function instances on a worker node. This extreme multi-tenancy requires strong isolation, so providers tend to deploy instances in dedicated VMs~\cite{agache:firecracker,web:gvisor,randazzo:kata}. Besides isolation, these general-purpose VMs provide seamless ecosystem compatibility for application developers, i.e., supporting popular libraries and SDKs along with the familiar POSIX interface. This deployment model, however, comes with non-negligible overheads towards the two key deployment constraints: CPU cycles and memory in the cloud fleet.

% Hence, production platforms commonly rely on general-purpose sandboxes such as Firecracker MicroVMs~\cite{agache:firecracker} or containers~\cite{randazzo:kata,web:gvisor}. 
% %\dmi{any multi-tenancy in production requires isolation, this reads like you only need virtualization if you collocate 100s of VMs; misleading}
% These sandboxes provide backwards compatibility through POSIX support, which is important for development simplicity.
% However, they come with non-negligible memory and CPU overheads that inherently limit deployment density.

% To understand these overheads, one must look at how existing serverless platforms work. 
In this paper, we ask the fundamental question: \emph{how can cloud providers achieve high deployment density while retaining ecosystem compatibility?} To answer that, we examine the root causes of resource inefficiencies in such environments. 
Since serverless functions are stateless, they transfer data between caller and callee functions via external remote storage services~\cite{klimovic:pocket, sreekanti:cloudburst,mvondo:ofc}.
% must communicate with remote storage to retrieve necessary data and persist outputs. 
To facilitate this securely, current architectures force every function instance to load and execute its networking stack, RPC libraries, and cloud service SDKs, which we refer to as the \emph{communication fabric}.
Hence, each instance couples application logic with the required I/O processing within its sandbox, leading to massive memory footprint duplication and CPU overhead from repeatedly crossing virtualization boundaries, substantially reducing deployment density.

To understand the key factors preventing higher deployment density, we break down CPU cycles and memory footprint on worker nodes in a serverless cluster across the application and virtualization stacks.
% characterize the two resources that constrain it, namely CPU and memory during the execution of modern serverless functions. % and investigate why they are limited. 
% \dmi{the below sentence comes out of the blue, drop}
% \sout{We find that 82\% of the most popular functions~\cite{awsserverless} on AWS Lambda rely on cloud provider SDKs, confirming the dependence of serverless functions on a communication fabric. This I/O-heavy communication fabric bloats both CPU and memory usage. }
% \dmi{this is too low-level and distracting, drop}
Our study of CPU cycle breakdown reveals that communication fabric execution often accounts for the largest fraction (74\%) on the worker nodes, exacerbated by virtualization and the inefficiency of the high-level language runtimes chosen by application developers who prioritize time-to-market.
% its execution within VMs.
% Our study of the CPU cycles distribution shows that 60\% of CPU 
% cycles belong to the execution of the communication SDK libraries, e.g., AWS S3 storage~\cite{aws_s3}, and high-level \dmi{function-invocation} protocol handling (e.g., HTTP or gRPC~\cite{google:grpc}), \dmi{which altogether we refer to as the serverless communication fabric.}
% from the CPU's point of view, user-space cycles represent the execution of functions, but a large fraction (60\%) is consumed by SDK library execution and high-level protocol handling (e.g., HTTP). 
As for the memory footprint on a worker node, the communication fabric accounts for over 25\% of a function's total memory footprint. 
Thus, this massive replication of the communication fabric across 100s of VMs colocated on each node results in gigabytes of memory occupied by duplicate code.

% % the coupled design and inherent problems to it (which is introduced in section 3)
% These costs are not merely implementation overheads. 
% \dmi{which costs? also: never start with a negative sentence, it's meaningless (i.e., This is not A; then what is this? content-free sentence}
% They stem from the current architectural boundary. Because application logic and provider-managed I/O are tightly coupled and live inside the same sandbox, the request path is tightly serialized: restore the VM, receive the invocation, fetch remote inputs, execute user logic, write results, and only then return. The same reason bloats snapshots and restore times,
% leaving the host without global visibility limiting efficient coordination of I/O demands.

% \dmi{the para is right but w/o a proper trad arch description, it doesn't read clearly}
We argue that these overheads are intrinsic to current architectures that tightly couple application logic with I/O processing within isolated sandboxes, thereby imposing additional penalties in serverless environments. This coupled design strictly serializes the execution critical path (sandbox init, fetch, compute, write) and inflates function initialization times due to bloated memory snapshots.
% \sout{, and leaves the host infrastructure without a global view of the node's network I/O traffic.}}
% \dmi{K, your fix here doesn't work: still too abstract and even misleading bc host does NOT manage IO at all in the baseline}
% \jovan{This paragraph tries to make three separate points (serialization, bloated snapshots, lack of global visibility) but mushes them together into two sentences. Each point deserves its own clear sentence. Also, ``The same reason bloats snapshots ...'' is a run-on sentence that is hard to parse.}
%leaving the host blind to application-level I/O intent and forcing conservative per-VM controls instead of flexible node-wide coordination. 

Previously proposed systems aim to mitigate the above issues, but often at the expense of compatibility, which is essential for customers of production serverless platforms. These systems tend to rely on WebAssembly~\cite{shillaker:faasm}, library OSes~\cite{li:jord,you:alloystack}, or single-address-space mechanisms~\cite{li:jord,fried:junction}, introducing disruptive changes into the programming and deployment models, requiring rewriting application code to use their API or manual decomposition of the computing and IO, as in Dandelion~\cite{kuchler:dandelion}. Disconnected from the rich Linux and popular libraries ecosystem, such solutions make code maintenance and support for popular runtimes extremely challenging~\cite{wasm-limit,wasm-gap,wasm-state}. Thus, cloud providers tend to prioritize ecosystem compatibility over lightweight hypervisors; for example, Google Cloud Run notably reverted from its custom lightweight sandbox, gVisor~\cite{web:gvisor}, back to a fully compatible KVM-based hypervisor in its second generation~\cite{cloudrun-gen2}.

To showcase that high deployment density can be achieved without compromising compatibility and performance under strict SLOs,
% , or SLO response-time attainment, 
we introduce \emph{\sys}, a serverless-native KVM-based hypervisor. \sys slashes the per-VM CPU and memory overheads of the communication fabric and virtualization stack while preserving full compatibility with the conventional FaaS programming model. \sys achieves this efficiency by fundamentally decoupling I/O processing from the application logic, transparently offloading I/O handling to a shared, highly concurrent backend service running natively on the host. In \sys, function instances still run in dedicated VMs but communicate via fully backward-compatible provider SDK frontend libraries. These thin frontends enable communication with the shared backend via API remoting over zero-copy shared memory~\cite{yu:ava,qi:palladium,kim:linefs}, removing the heavy networking stacks from the guest. 

\sys efficiently reuses CPU cycles and memory—previously occupied by the duplicated communication fabric—to host a greater number of co-resident function instances.
Furthermore, \sys's decoupled architecture enables several asynchronous optimizations that are incompatible with traditional, coupled designs. First, by leveraging deterministic routing hints injected by the platform's ingress layer, \sys completely overlaps input payload prefetching with VM bootstrapping. Second, \sys allows the function to finish processing the invocation before writing its output payloads back to remote storage; the host backend independently completes the background write while retaining at-least-once execution semantics. Crucially, \sys achieves this with zero user code modifications while hardening the node's threat model, as the cluster orchestrator provisions least-privilege identity tokens directly to the trusted backend, keeping raw provider credentials entirely out of the untrusted guest VM.

We prototype \sys by extending Firecracker~\cite{agache:firecracker} with a shared-memory communication transport—running atop TCP and RDMA—and a frontend library that transparently remotes the AWS S3 SDK API. We evaluate \sys deployed atop a Knative cluster using the vHive framework~\cite{ustiugov:reap} and a mix of compute- and I/O-intensive functions from the vSwarm benchmark suite~\cite{vswarm}. We show that \sys reduces node-level CPU and memory usage by up to 44\% and 31\%, respectively, yielding a 37\% improvement in deployment density under strict response-time SLOs, with RDMA accounting for 50\% of this gain. Furthermore, \sys reduces warm- and cold-start latencies by 39\% and 10\%, respectively, bringing response times within 20\% of those of an ecosystem-incompatible, WASM-based hypervisor, proving that extreme density and high performance do not require sacrificing legacy compatibility.

%% file: sec/2_background.tex
\section{Background on Serverless Clouds}
\label{sec:background}

\subsection{Programming Model \& Economy}
\label{sec:background_serverless}

% how to program and deploy with FaaS: convenience at the cost of offloaded management, hence deep stack and overheads

% providers host and colocate, making deployment density crucial for economical feasibility

% serverless cloud exist within a more broad cloud ecosystem with k8s and containers and VMs. Hence, back-compat with the above is key. Alternative solutions, such as specialized hypervisors and control planes, foregoe compatib, preventing its adoption in production. 

% Takeaway: is it possible to keep back-comp with high dep density?

% The advent of serverless computing, or Function-as-a-Service (FaaS), has fundamentally shifted how cloud applications are developed and deployed. 
In the serverless paradigm, developers focus on their applications, while deployment and resource management are handled by the cloud provider. Developers write business logic as stateless functions in high-level languages, such as Python and NodeJS~\cite{datadog}, use third-party libraries for processing, and connect them into application workflows. These functions typically rely on cloud SDKs to interact with remote storage and on RPC libraries to handle function invocations. Specifically, our analysis of 362 functions from the 50 most popular applications in the AWS Serverless Application Repository~\cite{awsserverless} shows that $82\%$ of these functions use cloud provider SDKs (AWS S3, ElastiCache, DynamoDB) to communicate across functions, making provider SDKs the de facto standard for I/O in serverless clouds.

To make this execution model economically viable, cloud providers must heavily amortize infrastructure costs by maximizing deployment density, by collocating hundreds to thousands of function instances on each worker node. This extreme multi-tenancy necessitates stringent security boundaries, lean sandboxes, and execution environment with minimal CPU and memory overheads.

% Another strict requirement for 
Also, serverless cloud programming and deployment models require seamless ecosystem compatibility. Existing applications are heavily anchored to high-level languages by domain-specific dependencies—such as Python's machine learning ecosystem and Node.js's extensive API SDKs. These dependencies introduce significant migration barriers because they lack the maturity of high-performance compiled runtimes like C++ or Rust. Consequently, preserving compatibility with existing FaaS programming models, containerized deployment strategies, and POSIX interfaces is imperative to minimize migration friction, reduce time-to-market, and simplify maintenance.

% Another strict requirement for serverless cloud programming and deployment models is backward compatibility, which is essential for applications that rely on specific popular libraries available for high-level languages, e.g., Python's \texttt{pandas} and \texttt{PyTorch} and JavaScript's API SDK glue are barely replaceable by alternatives in more efficient runtimes such as C++ and Rust.
% % and heavily reliant on popular libraries. 
% Compatibility with the FaaS programming model, container-based deployment, and the POSIX interface substantially reduces time-to-market and simplifies maintenance.

% \dmi{below should be shortened and incorporated into 2.2}
% Most providers use general-purpose MicroVMs~\cite{agache:firecracker,randazzo:kata,crosvm,azure-vm,cloud_hypervisor}
% that support the entire POSIX API, offering substantial compatibility for application developers, albeit at the cost of increased memory and CPU overhead, thereby affecting deployment density. Alternative solutions, such as specialized sandboxes~\cite{web:gvisor,shillaker:faasm, kuchler:dandelion} and library OSes~\cite{you:alloystack, li:jord,shen:xcontainers},
%\dmi{references need to be close to the specific examples, not bundled altogether}
% forego compatibility. They force developers to adopt a different programming model in favor of specific runtimes, change their deployment automation to take into account augmented/modified libraries thereby preventing their adoption in production.
% Hence, the key question is how to maintain backward compatibility while achieving higher deployment density.

\subsection{Today's Serverless Cloud Architecture}
\label{sec:background_serverlessarch}

\begin{figure}
    \centering
    \includegraphics[width=\linewidth]{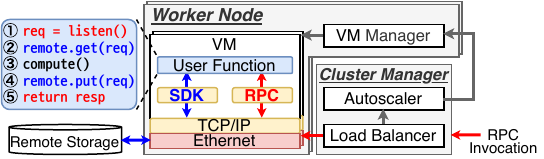}
    \caption{Traditional serverless architecture overview.
    % \TODO{K, shrink the fig vertically, it's huge}
    % \TODO{ETH to Ethernet}
    % \TODO{TCP/IP has to be as wide as Ethernet}
    % \TODO{Remove "Node Component" but leave VM Manager as is}
    % \TODO{low-prior: the fig and fonts are too large}
    }
    \label{fig:traditional_arch}
    \vspace{-17pt}
\end{figure}

% takeaway: there are two density constraints: CPU and memory. Need to characterize
Figure~\ref{fig:traditional_arch} shows a modern serverless platform, similar to AWS Lambda~\cite{agache:firecracker} and Google Cloud Run~\cite{googlecloud,knative}, comprising a cluster manager that handles incoming function invocations via its Load Balancer, which routes HTTP API invocations (relying on an underlying RPC stack) to active function instances right away or after requesting new instances from the autoscaler. The autoscaler monitors instance load and adjusts the number of instances by sending commands to the VM manager that creates VM instances and configures their CPU and memory quotas.

% Within this environment, a serverless function \kev{follows a strict sequential execution flow.}
% \dmi{the first sent says: A is not xxx, then what is A? this only distracts the narrative}
% \dmi{what's strictly ephemeral execution flow? what's ephemeral about it? why do we need to introduce a new term?}
% As detailed in the VM block of Figure~\ref{fig:traditional_arch}, 

Figure~\ref{fig:traditional_arch} show the function invocation lifecycle that consists of five steps: The invocation first arrives at the load balancer that forwards it to a function instance, which listens for it with its RPC interface \circled{1}, such as gRPC~\cite{google:grpc}, hosted on a worker node.
% The function instance waits for an invocation request routed by the load balancer through the RPC stack. 
\circled{2} Upon receiving a request, the instance starts processing the invocation, typically followed by using the cloud SDK API to fetch required inputs from remote storage (e.g., AWS S3, DynamoDB, ElastiCache, Azure Blob Storage~\cite{aws_s3,aws_elasticache,azure_blob}) over HTTP. \circled{3} Then, the user code performs its core computation logic. \circled{4} It then stores the resulting data back to the remote storage via the provider SDKs. \circled{5} Finally, the instance returns a response to the invocation caller through the same RPC interface, before moving to processing the next invocation.

Each instance encapsulates a fully virtualized stack running an HTTP server with a user-defined handler that operates atop of the \emph{communication fabric} that comprises the RPC protocol used by the invocations and a variety of provider SDKs necessary for communication with storage and cache services, which operate atop the TCP/IP in the guest OS and virtio devices emulated by the hypervisor. 

Most providers use general-purpose MicroVMs~\cite{agache:firecracker,randazzo:kata,crosvm,azure-vm,cloud_hypervisor} that support the entire POSIX API, offering substantial compatibility for application developers, albeit at the cost of increased memory and CPU overhead, thereby reducing deployment density. 
% The cluster manager's autoscaler monitors workload demand and adjusts the number of function instances on each node.
Providers disable guest memory sharing among the co-resident VMs to prevent timing attacks~\cite{agache:firecracker,ustiugov:reap,firecracker-recomm}, causing substantial memory duplication and extra CPU overheads inherent in the virtualization and communication fabric stack, which may significantly reduce the overall deployment density, elevating today's serverless cloud's operational costs.

Next, we analyze the implications of this serverless architecture on the overall CPU and memory resource usage, and identify the key factors that limit the deployment density. 

%% file: sec/3_characterization.tex
\section{Quantifying Deployment Density Limits}
\label{sec:motivation}

We quantify the compute (\S\ref{sec:motivation_compute}) and memory (\S\ref{sec:motivation_memory}) overheads of the serverless communication fabric and virtualization stack, analyzing their root causes and why prior alternatives fail. Our study evaluates vSwarm~\cite{vswarm} functions on Knative/Firecracker~\cite{agache:firecracker}, overcommitting 280 VMs per worker node to match prior setups (details in \S\ref{sec:method}).

\subsection{CPU Overheads} 
\label{sec:motivation_compute}

% \cz{where is the virtualization tax in \S\ref{sec:background}?} DMI last para in 2.2
We analyze the CPU overheads limiting deployment density by decomposing worker-node cycles into three components: aggregate usage, intrinsic cloud I/O stack overhead, and virtualization overhead~(\S\ref{sec:background}).

\subsubsection{Worker Node Cycle Distribution.} 
\begin{figure}
    \centering
    \setlength{\abovecaptionskip}{0.05cm}
    \captionsetup[subfigure]{skip=2pt}

    \begin{subfigure}[t]{\linewidth}
        \centering
        \includegraphics[width=\linewidth]{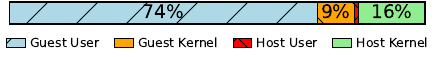}
        \caption{CPU cycles distribution on a worker node.}
        \label{fig:cpu-breakdown}
    \end{subfigure}

     \begin{subfigure}[t]{0.46\linewidth}
        \centering
        \includegraphics[width=\linewidth]{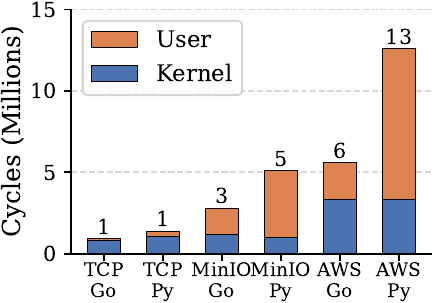}
        \caption{CPU cycles on a synthetic PUT workload.}
        \label{fig:runtime-cycles}
    \end{subfigure}\hfill
    \begin{subfigure}[t]{0.46\linewidth}
        \centering
        \includegraphics[width=\linewidth]{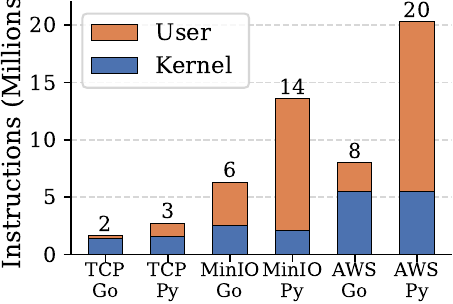}
        \caption{CPU instructions on a synthetic PUT workload.}
        \label{fig:runtime-instr}
    \end{subfigure}

    \begin{subfigure}[t]{\linewidth}
        \centering
        \includegraphics[width=.9\linewidth]{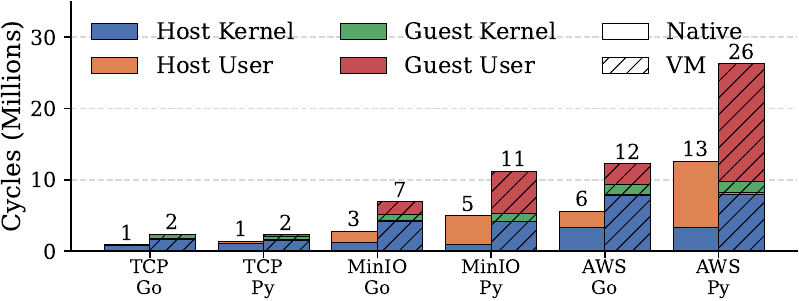}
        \caption{CPU cycles, bare-metal vs. virtualized for Go and Python.}
        \label{fig:virt-overhead}
    \end{subfigure}

    \caption{CPU cycles breakdown: (a) overall on a worker node across host and guest domains; (b) CPU cycles and (c) instructions for a synthetic single-PUT workload across different communication fabrics (TCP vs. SDKs) and (d) CPU cycles across bare-metal and virtualized environments.}
    \label{fig:cpu-overhead}
    \vspace{-12pt}
\end{figure}

We first study the aggregate CPU cycle distribution on a worker node running instances of a representative, balanced mix of 10 vSwarm functions. The load generator is configured so that each function contributes equally to CPU utilization\cz{cycles?}. Figure~\ref{fig:cpu-breakdown} shows that the guest user space constitutes the largest fraction of CPU cycles (74\%), while a substantial 25\% is spent in the kernel space, split between the host (16\%) and guest kernel space (9\%). In today's serverless architecture, the guest-user fraction includes both the user handler and the communication fabric, which incur overhead for constructing storage requests, marshaling data, managing connections, and executing the cloud I/O stack.
Also, the guest-kernel and host-kernel cycles are not application logic either; they are the cost of driving that I/O through the virtualized network stack. 
To pinpoint the exact overheads, we further break down these layers with a microbenchmark.

\subsubsection{Decomposing Transport Cost from SDK Cost}
\label{sec:motiv_transport-cost}

% To separate the overhead of the communication fabric from application logic, we profile a synthetic single-PUT microbenchmark that writes a 1MB payload to MinIO~\cite {minio}, the leading open-source alternative used in production Kubernetes clouds and fully compliant with AWS S3. We capture CPU cycles using \texttt{perf} while executing this operation in a tight loop over 2,000 iterations. 

To isolate communication fabric overhead from application logic, we profile a synthetic benchmark that performs a 1MB PUT to MinIO~\cite{minio} (a production-ready, S3-compliant datastore) using \texttt{perf}. We compare a baseline TCP socket-representing the minimum software cost for network transfer-against the MinIO and AWS S3 SDKs in Python and Go~\cite{datadog}.
% We first run this workload in a native environment and compare three implementations: a TCP socket, the MinIO S3-compliant SDK, and the AWS S3 SDK. We repeat this experiment with Go and Python SDKs, which are among the most popular runtimes in serverless clouds~\cite{datadog}, to assess the impact of the language runtime chosen by serverless application programmers. We include TCP as a lower bound on the minimum software cost of moving data over a commodity network, which, when compared to the SDK-based transfers, quantifies the added compute cost of the cloud-storage stack.

Figures~\ref{fig:runtime-cycles} show that the cloud I/O stack is inherently compute-intensive, driven largely by user-space tasks like request construction, serialization, authentication, and connection management. Compared to TCP, the MinIO SDK increases CPU cycles by 3x and 5x (for Python and Go, respectively), which we attribute to the increase in the number of executed instructions by 3x and 4.5x, respectively, due to the SDK overhead, as shown in Figure~\ref{fig:runtime-instr}. The AWS SDK similarly inflates cycles by 6x and 13x, which correlates with a similar increase in the instruction count. Crucially, this I/O overhead is coupled to the user's language choice, with Python being less efficient than Go~\cite{datadog}. Such architecture binds user logic and I/O processing within the same VM; thus, providers have to inherit the user's runtime inefficiencies, making it impossible to independently offload I/O to a more efficient language.
% \cz{Actually, it should evaluate the impact of the binding, but we don't have time.}

% First, Figure~\ref{fig:runtime-cycles} shows that the cloud I/O stack is compute-intensive even when running natively on a host. MinIO SDK uses $3\times$ and $5\times$, and the AWS SDK uses $6\times$ and $13\times$ more cycles than TCP for Python and Go, respectively. The same trend appears in Figure~\ref{fig:runtime-instr} about instructions count: MinIO SDK uses $3\times$ and $4.5\times$, and the AWS SDK uses $4\times$ and $10\times$ more instructions. Much of this growth occurs in user space, showing that the cloud I/O stack itself---request construction, serialization, authentication, buffering, and connection management---is a major source of compute overhead.
% Second, the overhead of the cloud I/O SDK is tightly coupled to the user’s language choice. Python-based SDK execution, one of the most popular in serverless environments \cite{datadog}, is markedly more expensive than Go, especially once the full cloud SDK is involved. This coupling is important: in the current architecture, the provider’s I/O path executes inside the same VM and runtime as the user function. As a result, the efficiency of the cloud I/O stack inherits the language chosen by the user. A provider cannot, for example, keep the user-facing function in Python while offloading remote I/O to a more efficient implementation, because the current architecture binds both concerns within the same isolated VM.

\subsubsection{The Amplification of Virtualization}
\label{sec_char_virt}

Next, we examine the CPU cycle breakdown for the same I/O path running inside a Firecracker VM, using the same MinIO microbenchmark. The goal is to see how much sandboxed execution within a VM impacts the I/O path. Figure~\ref{fig:virt-overhead} compares native execution with VM execution for the same single-PUT workload. 
Across all configurations, virtualization roughly doubles the total cycle overhead. 
We attribute this to the communication fabric running in the VM, which causes cross-boundary operations and triggers many KVM exits, which we study in detail in \S\ref{sec:eval_cpu_cycles}.
The communication fabric must route its network packets through the guest kernel network stack, virtualized network devices, and the host kernel network stack. Thus, these intermediate layers require CPU cycles, which is overhead that steals CPU time that would have been allocated to the actual functions' logic.

In summary, the heavy communication fabric inflates user-space cycles, while the virtualized network stack amplifies the kernel-space cycles due to the hypervisor activity.

\subsection{Memory Overheads}
\label{sec:motivation_memory}

\begin{figure}
    \centering
    \includegraphics[width=\linewidth]{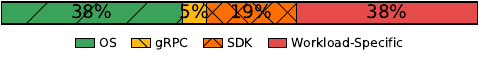}
    \caption{Breakdown of memory footprint for each component during function execution averaged across vSwarm workloads.}
    \label{fig:mem-footprint}
    \vspace{-20pt}
\end{figure}

Beyond CPU cycles, serverless infrastructure inflates VM memory footprint, limiting deployment density. We quantify this using vSwarm workloads (\S\ref{sec:method}) reading/writing to MinIO, deriving \texttt{VmRSS} from \texttt{/proc/(pid)/smaps}. To isolate components additively, we measure: (1) a Hello World function over vsock~\cite{hung2023vsock} (guest-OS baseline), (2) the same function using gRPC over TCP (RPC overhead), (3) adding the AWS S3 SDK for GET/PUT operations (SDK overhead), and (4) full vSwarm workloads.

Figure~\ref{fig:mem-footprint} reveals that the Cloud SDK ($19\%$) and RPC library ($5\%$) consume over $25\%$ of a function's total memory footprint. 
Serverless providers disable guest memory page sharing across VMs to prevent timing attacks~\cite{zhao:everywhere, deutsch:dagguise},
% \todo{find ref, see REAP}
resulting in this heavyweight communication fabric being duplicated within every isolated VM. A node hosting hundreds of instances~\cite{agache:firecracker} wastes gigabytes of physical memory, severely restricting multi-tenant density and inflating provider costs.

\subsection{Why Coupled Design is Ill-Suited for Serverless?}
\label{sec:motivation_sota}

The above compute and memory overheads are not merely implementation artifacts; they are intrinsic to the current serverless architecture, which tightly couples application logic with I/O processing and the communication fabric within isolated sandboxes. This coupled design inherently limits deployment density and induces serialization delays:

\noindent\textbf{Inflated Restoration Times.} 
To mitigate cold starts, providers increasingly rely on snapshot-and-restore mechanisms. However, because the memory footprint is bloated by heavy, I/O-centric SDKs and replicated RPC stacks (Figure~\ref{fig:mem-footprint}), the snapshots are excessively large. Reading these bloated images from disk and restoring them to memory significantly prolongs the time to restore function, thereby defeating the purpose of rapid scaling.

\noindent\textbf{Strict Execution Serialization.} 
The tight coupling of compute and I/O forces the function's lifecycle onto a strictly serialized critical path. An invocation must sequentially: restore the snapshot, fetch the payload from remote storage, execute the user logic, and write the results back. In this coupled design, the function's code drives its own I/O, which cannot start before VM bootstrapping finishes.

\subsection{Can Alternative Solutions Help?}
\label{sec:back_alternatives}

Prior works long ago identified the overheads of virtualization in the context of memory virtualization~\cite{margaritov:ptemagnet, ustiugov:reap}, bootstrapping time~\cite{du:catalyzer, liu:fastiov}, and I/O processing~\cite{guo:vpri}. However, most of them focused on reducing the CPU and memory overheads of the coupled designs rather than on a clean-slate solution.
Many works propose swapping conventional, KVM-based virtualized sandboxes in favor of specialized environments: 
library operating systems~\cite{wanninger:virtines, fried:junction, you:alloystack} and single-address-space~\cite{li:jord,kotni:faastlane, shillaker:faasm} mechanisms.
While these lightweight sandboxes reduce CPU and memory overhead, they forego backward compatibility with the FaaS programming model and POSIX API required for seamless usage of high-level language runtimes popular in serverless~\cite{datadog} and a wide range of publicly available libraries and modules.

Recognizing that the tight coupling of compute and I/O limits efficiency inherently, a recent system, Dandelion~\cite{kuchler:dandelion}, has structurally separated these domains. In particular, Dandelion explicitly separates computation from I/O, but requires manual application rewriting with a new API and, often, in a different language because maintaining the popular interpreted and JIT-ed runtimes is notoriously challenging~\cite{wasm-limit}.
Thus, in practice, serverless application programmers prioritize time-to-market over potential efficiency gains over pursuing efficiency goals at the cost of losing compatibility with the developer ecosystem, i.e., the wide variety of libraries, modules, and base images available for high-level languages, such as Python for machine learning. An illustrative example is the fate of gVisor~\cite{web:gvisor}, a unikernel-like hypervisor that Google used in the first generation of Cloud Run but subsequently reverted to a KVM-based hypervisor in the second generation~\cite{cloudrun-gen2}. 

In contrast, an ideal architecture must ensure high deployment density without disrupting the developer experience.

%% file: sec/4_design.tex
\section{\sys Design}
\label{sec:design}

Given the insights from \S\ref{sec:motivation}, we introduce \emph{\sys}, a serverless-native I/O hypervisor that fundamentally rethinks function execution. We build \sys with three main ideas.

First, \sys \emph{decouples I/O from compute}, by offloading I/O handling from each VM to a separate execution domain.
\sys separates user computation from provider-managed I/O and executes the latter in a shared node-local backend. This removes the duplicated infrastructure stack from the common path, reducing the compute and memory overhead (\S\ref{sec:motivation_compute}--\S\ref{sec:motivation_memory}), and the inflated restoration time caused by bloated VM state (\S\ref{sec:motivation_sota}). 
At the same time, \sys preserves ecosystem compatibility by keeping the user-visible invocation and service APIs unchanged and by retaining a legacy path for uncommon networking behaviors.

Second, \sys makes \emph{I/O asynchronous with respect to VM execution}.
Once I/O is no longer tied to the lifetime of a single VM, \sys can overlap remote fetches with VM creation and allow remote writes to complete after the function invocation's processing completes. 
This breaks the strict restore--fetch--compute--write serialization, identified in \S\ref{sec:motivation_sota}, and shortens the critical path of the invocation. 
\sys does so while preserving safety: the backend takes responsibility for performing the I/O transfer, e.g., to remote storage, on behalf of the function instance, which can then proceed to execute the next incoming invocation. 

\subsection{Architecture and Abstractions}
\label{sec:design_overview}

\begin{figure}
    \centering
    \includegraphics[width=0.95\linewidth]{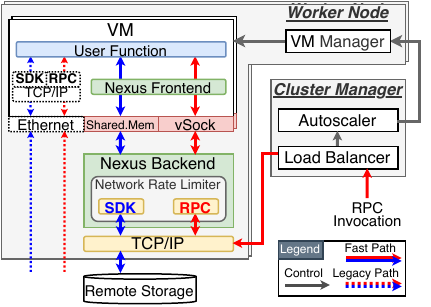}
    \caption{\sys serverless architecture overview.
    }
    \label{fig:proposed_arch}
    \vspace{-12pt}
\end{figure}

As illustrated in Figure~\ref{fig:proposed_arch}, \sys fundamentally reshapes the serverless worker node while leaving the overarching cluster control plane—comprising the load balancer and autoscaler—entirely unmodified. On the worker node, the architecture is split into two distinct execution domains: a lightweight \sys frontend residing within each isolated tenant virtual machine, and a trusted, highly concurrent \sys backend operating natively on the host. 

The core of the \sys design is establishing the remoting boundary at the high-level application programming interfaces of cloud service SDKs and function invocation RPCs. Instead of executing this heavyweight communication fabric inside the guest, the user's function interacts with the thin \sys frontend, which seamlessly forwards these operations to the \sys backend. The backend, acting as a shared data plane for all co-resident VMs, encapsulates the network rate limiter, the full SDK logic, and the transmission control protocol stack. This transparent offloading successfully amortizes the infrastructure tax across the host without violating the expected semantics of the conventional serverless programming model.

To ensure strict POSIX compliance and support for arbitrary workloads, the architecture defines a bifurcated network flow comprising a fast path and a legacy path. Compliant cloud invocations and managed storage requests travel over the optimized, low-latency fast path, using low-latency virtual sockets for control messages and zero-copy shared memory for bulk data transfers between the frontend and backend. Conversely, if a function bypasses the provider interfaces to perform low-level networking, it triggers the legacy path, which transparently falls back to standard virtualized Ethernet devices governed by the same fixed-rate-limiting mechanisms as the baseline architecture.

%%%%%%%%%%%%%%%%%%%%%%%%%%%%%%%%%%%%%%%%%%%%

\subsection{Anatomy of an Invocation}
\label{sec:anatomy_invoc}

\begin{figure}
    \centering
    \includegraphics[width=\linewidth]{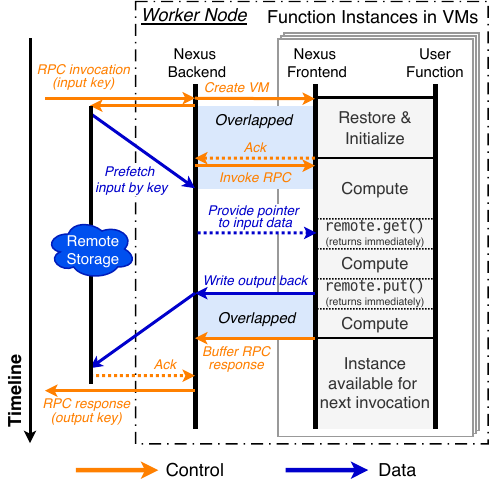}
    \caption{Function execution lifecycle with RPC management and cloud storage access offloading.
    }
    \label{fig:design_rpc_s3}
    \vspace{-15pt}
\end{figure}

This decoupled architecture fundamentally transforms the traditionally serialized serverless lifecycle into a highly pipelined and asynchronous execution model, as depicted in Figure~\ref{fig:design_rpc_s3}. In the baseline coupled architecture~(\S\ref{sec:background_serverlessarch}), a serverless platform must strictly serialize the VM restoration, runtime initialization, fetching of remote inputs, and the execution of user logic. By shifting the invocation termination to the host backend, \sys effectively hides network latency from the VM's critical path and unlocks asynchronous optimizations that significantly speed up cold and warm invocations (\S\ref{sec:eval_ablation}).
% reduce both cold- and warm-start delays (\S\ref{sec:eval_ablation}).

\subsubsection{Invocation Interception and Parallel Provisioning}
When a new request arrives at a worker node, the shared \sys backend acts as the authoritative first recipient, completely shielding the guest environment from the initial network transaction. Instead of routing incoming network packets through the host bridge and into the guest operating system's network stack, the backend terminates the RPC connection natively on behalf of the function instance. This early interception is a critical departure from existing architectures, as it grants the host infrastructure immediate visibility into the request payload before the function instance's VM is ready.

Because the backend fully owns this early lifecycle phase, it can instantly evaluate the request metadata and orchestrate the necessary provisioning in parallel. Upon unpacking the RPC, the backend simultaneously triggers the host's VM manager to begin restoring a VM from its snapshot on disk. This eliminates the baseline inefficiency in which the RPC server cannot even begin accepting connections until the entire VM and guest runtime have fully booted and initialized.

\subsubsection{Asynchronous Input Prefetching}

To overlap network communication with compute provisioning, \sys capitalizes on the predictable nature of serverless data dependencies. Our manual analysis of 362 functions from the 50 most popular applications in the AWS Serverless Application Repository~\cite{awsserverless} shows that $96\%$ of functions have deterministic inputs known at invocation time. 
Crucially, extracting these hints requires zero modifications to the user's application code. Modern serverless orchestration frameworks and event sources (e.g., AWS API Gateway, Step Functions, or Knative Eventing) inherently parse incoming event payloads to route requests. \sys leverages this existing platform infrastructure by having the cluster's ingress layer automatically promote known data dependencies—such as target S3 bucket and key names found in the JSON trigger event—directly into the RPC metadata headers before the invocation ever reaches the worker node. 

The \sys backend parses these embedded hints to completely overlap the remote input fetching with the VM's bootstrap phase. Using the provider's managed credentials for that specific function, the backend immediately authenticates and initiates the remote storage \texttt{GET} operation. By the time the VM is fully restored and the user handler is invoked, the input payload is either actively streaming or already fully downloaded, effectively masking the network delay from the guest's execution timeline.

Furthermore, this prefetching mechanism is tightly integrated with the system's memory management. The backend uses the payload-size metadata provided in the invocation hints to precisely allocate a dedicated shared memory region tailored to the incoming object's dimensions. This guarantees optimal memory utilization on the host and ensures that the guest environment does not need to dynamically resize buffers or handle complex memory allocations during the critical path of its execution.

\subsubsection{Streaming Fallback for Opaque Payloads}
While hint-based prefetching covers most standard serverless workflows, \sys must robustly handle scenarios in which input sources are entirely dynamic. For the minority of functions where input hints cannot be determined prior to execution (a mere 4\% of the 362 functions in the AWS repository~\cite{awsserverless}), 
or where the payload size is completely opaque to the caller, the system cannot safely preemptively map a perfectly sized shared memory region. In these edge cases, \sys safely defaults to synchronous data retrieval using fixed-size circular buffers established between the frontend and the backend.

This streaming fallback mechanism guarantees correct execution and strictly bounds memory consumption for arbitrary workloads, preventing memory exhaustion attacks or faults caused by unexpectedly large payloads. The frontend continuously pulls chunks of data through the circular buffer as the user function consumes the input stream. 
While this approach is highly resilient, it inherently sacrifices the latency benefits of overlapped network transfers because the payload dimensions cannot be preemptively mapped and fetched during the VM boot phase (\S\ref{sec:eval_cpu_cycles}).

\subsubsection{Transparent I/O Remoting During Compute}
Once the VM is fully initialized and the user handler begins executing its core logic, it issues requests to retrieve its required data. In a traditional, coupled architecture, calling a cloud storage SDK triggers a cascading sequence of complex operations: constructing an HTTP request, establishing a secure socket layer connection, and pushing packets through the heavily layered guest and host network stacks. In the \sys architecture, these calls bypass the traditional guest networking stack entirely.

The \sys frontend acts as a lightweight interception stub. When the user code issues a standard SDK call, the frontend merely traps this request at the API boundary. Because the backend has already prefetched the necessary data based on the initial RPC hints, no actual network transmission occurs during this phase. The frontend simply immediately returns a pointer to the data residing in the strictly pre-allocated shared memory region pre-populated with the retrieved input data. This dramatically reduces the number of CPU cycles consumed by the guest and eliminates the virtualization overhead typically associated with heavy I/O processing (\S\ref{fig:cpu-breakdown}).

\subsubsection{Asynchronous Output and Early VM Release}
The final bottleneck in a coupled serverless architecture occurs during the teardown phase. Functions typically conclude by issuing a remote \texttt{PUT} operation to persist their outputs to a cloud storage bucket. In the baseline system, the VM compute resources are held captive, sitting completely idle while waiting for the remote storage service to process the write and return a network acknowledgment. \sys introduces an opt-in optimization that makes these remote writes fully asynchronous, drastically increasing deployment density by freeing compute resources sooner.

% \cz{Does a function write or read the remote repo only once during the execution? Better to mention it in the background.}
When the function completes its computation and issues a remote write, the frontend delegates the payload directly to the backend and immediately returns control to the function runtime. The function safely terminates its execution phase, allowing the worker node to immediately recycle or release the VM compute resources for subsequent warm invocations. The backend, now holding the output payload, independently drives the network write to completion in the background without tying up a dedicated VM.

Crucially, this aggressive early-release mechanism does not compromise the platform's strict consistency guarantees. To perfectly preserve the at-least-once execution semantics expected by serverless developers~\cite{fox:harvest, lee:linearizability}, \sys buffers the function's final RPC execution response. The backend only releases this final success response back to the caller after the remote storage layer explicitly acknowledges the successful write operation. If the background write fails, the backend accurately propagates the error, ensuring the caller never observes a successful execution for before the data has been persisted.

\subsection{Control and Data Plane Mechanisms}
\label{sec:control_data_planes}

To ensure that crossing the virtualization boundary does not introduce prohibitive latency that would negate the benefits of offloading, \sys completely circumvents standard virtual network devices. Instead, it employs a highly specialized, dual-channel transport design that distinctly separates orchestration traffic from bulk object payload transfers.

\subsubsection{Control and Data Plane Separation}
\sys splits communication between the frontend and the host backend strictly based on payload size and latency requirements. Lightweight control messages, RPC invocation metadata, and small SDK API requests require microsecond-scale responsiveness. To accommodate this, \sys routes the control plane over a low-overhead host-guest socket connection. Within our AWS Firecracker prototype~\cite{agache:firecracker}, this is implemented by exposing \texttt{virtio-vsock} within the guest VM. The hypervisor then binds this interface to a Unix Domain Socket on the host, providing a highly reliable, low-latency channel for the backend to consume and govern execution.

Conversely, bulk data payloads moving to and from remote cloud storage must avoid the severe CPU penalties associated with socket-buffer copying and kernel network stack traversals. \sys routes these large transfers through a dedicated data plane built entirely on zero-copy shared memory. This is implemented utilizing file-backed memory initialized with the \texttt{MAP\_SHARED} flag, which Firecracker subsequently surfaces to the guest operating system as an emulated peripheral component interconnect device. By mapping this region directly into both the guest and host address spaces, the frontend and backend can exchange gigabytes of payload data without a single memory copy.

\subsubsection{SDK Remoting Implementation}
The API remoting logic bridging these two planes consists of a deliberately thin interception library within the guest VM. This frontend stub cleanly mirrors the standard AWS Python Boto3 SDK and gRPC interfaces, ensuring that user applications require absolutely zero code modifications. When a function invokes a storage method, the frontend simply marshals the request parameters and pushes them across the control socket, leaving the heavy lifting of connection pooling, cryptographic signing, and HTTP request formatting to the host.

% \jy{Operating on the host side, the highly concurrent 
We implement the \sys backend with 7827 Golang LoC and the frontend with 645 Python LoC, given Python's dominance in serverless clouds~\cite{datadog}. The frontend is compatible with the AWS boto3 S3 GET/PUT API.
% ,the workloads' language—and also consists of 1KLoC}. 
Using Go for the \sys backend balances extreme concurrency with highly efficient memory and CPU utilization, allowing a single backend process to effortlessly multiplex I/O for hundreds of co-resident VMs. Furthermore, because the backend directly controls the physical networking stack, it is entirely free from guest operating system constraints. 

% This architectural liberation allows the b
\sys{}'s decoupled architecture enables seamless support for multiple network types. Specifically, commodity hosts can run the \sys backend over TCP, whereas more advanced setups can run \sys over an RDMA network, supporting kernel-bypassed remote direct memory access for data transfers (\S\ref{sec:eval_cpu_cycles}) -- transparently to applications. When a \sys backend retrieves an object via RDMA, the physical network interface card places the payload directly into the shared memory region, bypassing both the host and guest kernels. When operating on legacy hardware or communicating with storage endpoints lacking RDMA capabilities, the backend gracefully and transparently falls back to TCP. 
% without altering the frontend's execution state.

\subsubsection{Security and Isolation of Shared Memory}
Consolidating I/O operations within a shared host component necessitates uncompromising security guarantees to satisfy production cloud requirements. \sys maintains extreme multi-tenant isolation by strictly enforcing that memory is never globally accessible across co-resident VMs. The system provisions a dedicated, one-to-one mapping of an isolated shared memory region exclusively between a single tenant's frontend and the trusted host backend. There is no peer-to-peer mapping; thus, a compromised VM cannot read, write, or even address the data plane of a neighboring function.

Furthermore, the \sys backend itself operates entirely within the cloud provider's trusted host environment and is written in a memory-safe language, structurally preventing standard buffer overflow attacks from leaking cross-tenant data. For defense-in-depth deployments, cloud providers can further lock down these dedicated memory mappings using hardware-assisted memory protection extensions, such as Intel MPK~\cite{intel-mpk} or Arm CHERI~\cite{watson:cheri}, as demonstrated by prior works~\cite{kuchler:dandelion,fried:junction}. These hardware constraints ensure that even if the backend is compromised, unauthorized memory access remains physically isolated at the silicon level.

Beyond memory isolation, \sys fundamentally hardens the serverless threat model through centralized, least-privilege credential management. In a traditional architecture, raw provider credentials (e.g., AWS secret access keys) must be injected directly into the untrusted guest VM to enable SDK operations, creating a severe vulnerability in the event of arbitrary code execution or a sandbox escape. \sys completely eliminates this attack vector. The cluster orchestrator provisions short-lived, least-privilege identity and access management (IAM) tokens specifically bound to each function sandbox, securely supplying them exclusively to the trusted host backend. Because the \sys backend authenticates and fetches remote objects on behalf of the function, the raw cryptographic keys are never exposed to the user's execution environment, drastically reducing the blast radius of a compromised workload.

\subsection{Resource Management and Billing}
\label{sec:design_fairness}

\sys resource management operates similarly to the baseline design, where each VM runs in a cgroup, and each virtio-thread is limited to the fixed transmission rate, e.g., at 600Mbps, similar to AWS Lambda\cite{awslambda-network-bw}.
We implement a similar rate-limiting mechanism in the \sys backend using \texttt{golang.org/x/time/rate} for each SDK client. If a function instance requires several clients, e.g., to communicate with AWS S3 and DynamoDB, the rate limit is divided equally for each client. In our experiments, we observe little sensitivity to the transmission rate above 600 Mbps for the function mix we use for evaluation, which includes both compute- and I/O-intensive functions.

\section{Discussion and Limitations}

Consolidating I/O processing into a shared host backend inherently widens the cross-tenant fault domain. \sys{} mitigates this via a memory-safe implementation and a stateless, crash-only design: if the daemon faults, a host supervisor rapidly restarts it while frontend stubs transparently retry requests, converting potential failures into transient latency spikes. For stricter security, production deployments could further enforce silicon-level isolation using hardware memory protection extensions (e.g., Intel MPK, CHERI). 

Furthermore, while kernel-bypassing RDMA maximizes our peak deployment density gains (37\%), the architectural decoupling alone yields an 18\% improvement over standard TCP. This confirms that \sys{}'s structural separation provides fundamental resource efficiency even on commodity network hardware. \cz{a new paragraph} Finally, although prototyped for Python workloads, extending \sys{} to other prevalent FaaS runtimes (e.g., Node.js, Java) relies on a deliberately thin frontend interception stub ($\sim$600 LoC). This avoids the complex, low-level runtime modifications typical of ecosystem-incompatible sandboxes, preserving the developer experience across languages.

%% file: sec/6_evaluation.tex
\section{Methodology}
\label{sec:method}

\textbf{Hardware and software setup.} For all experiments, we use a 10-node c6620 CloudLab cluster. Each node has a 28-core Intel Xeon Gold 5512U CPU fixed at 2.1 GHz, 128 GB of DRAM, and a 100 Gbps Intel E810-XXV NIC. We run vHive~\cite{ustiugov:reap} running Knative~\cite{knative} v1.13 on top of Kubernetes~\cite{k8s} v1.29, and use Firecracker~\cite{agache:firecracker} v1.14 hypervisor for isolating as function instances. Upon cold starts, the system restores function instances running in Firecracker VMs from a snapshot with REAP~\cite{ustiugov:reap}, the technology that pre-records and inserts the functions' working sets into the VMs to minimize page faults. The guest OS is Linux v6.1 with Ubuntu 24.02. 
We deploy one master node, one load-generator node, 4 worker nodes, and 4 nodes for remote storage to make sure storage is never a bottleneck in our setup. The storage nodes run MinIO~\cite{minio}, a widely used open-source distributed storage service used in industry, behind Istio~\cite{istio}, and serve as the object store for the data path.

\textbf{Workloads} 
We use ten Python functions from the vSwarm~\cite{vswarm} suite, ordered from the most I/O-intensive to the most compute-intensive: stack training's reducer (ST-R), lightweight ML inference (LR-S), encryption (AES), web serving (WEB), stack training's trainer (ST-T), RNN serving (RNN), JSON deserialization (MAP, RED), CNN Serving (CNN), and image resize (IR).
These workloads encompass a broad spectrum of compute- and I/O-intensive functions, with compute-to-I/O execution time ratios ranging from 10\% to 90\%, effectively representing serverless behavior~\cite{romero:faast}. To drive representative arrival patterns, we use In-Vitro~\cite{ustiugov:invitro}, which plays sampled Azure Function traces~\cite{shahrad:serverless,azure_trace}.
We sample these traces so that CPU utilization for each workload type, e.g., web serving and map-reduce, stays the same. We run the trace for 32 minutes, including a 2-minute warm-up period. After warmup, we introduce 20 new functions (2 sets of workload suites), increasing CPU load by 5\% across the cluster at each load step.

\textbf{Deployment density and other metrics.}
We define deployment density, our key optimization metric, as the maximum number of user functions a cluster can serve while satisfying the target SLO (p99 latency < 5 $\times$ unloaded latency calculated for each function individually). Deployment density can also be considered the throughput of a serverless system, since each function deployment incurs a series of invocations, as shown in the sampled trace.
We also evaluate the system's CPU and memory footprint as key deployment-density constraints, along with warm and cold response times.

\textbf{Systems, variants, and comparison scope.} 
We compare four configurations. 
The first configuration, \textit{Baseline}, illustrates the current paradigm of VM-based serverless computing, maintaining both the gRPC server and the Boto3 SDK within the VM environment. Next is \textit{\sys-TCP}, which offloads provider SDK operations and streamlines the invocation RPC path. The third, \textit{\sys-Async}, implements input prefetching and the early release of VMs for remote write operations on top of \sys-TCP. 
Finally, we have \textit{\sys}, which replaces TCP transport with RDMA. 
We also compare against \textit{Faasm}~\cite{shillaker:faasm}, a state-of-the-art for WebAssembly-based hypervisor that foregoes compatibility with the programming model and image ecosystem.

\section{Evaluation}
\label{sec:eval}

In this section, we evaluate the design and implementation of \sys.
We first evaluate whether \sys improves deployment density in a cluster with a realistic mix of functions (\S\ref{sec:eval_e2e}), and then explain the resulting gains through an ablation-driven analysis of warm-path CPU cycles, memory footprint, and cold-start latency (\S\ref{sec:eval_ablation}). We then compare \sys against Faasm, a WebAssembly-based lightweight hypervisor, in a focused case study to gauge the remaining efficiency gap to a lightweight but ecosystem-incompatible runtime.

\subsection{End-to-End Evaluation}
\label{sec:eval_e2e}

\begin{figure}
    \centering
    \setlength{\abovecaptionskip}{0.05cm}
    \captionsetup[subfigure]{skip=2pt}

    \begin{subfigure}[t]{\linewidth}
        \centering
        \includegraphics[width=\linewidth]{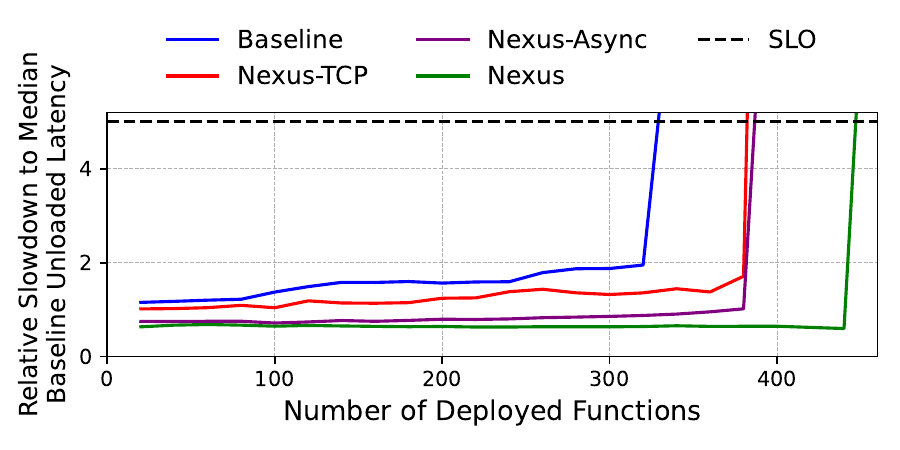}
        \caption{End-to-End Latency.}
        \label{fig:mix-trace-e2e}
    \end{subfigure}
    
    \begin{subfigure}[t]{0.49\linewidth}
        \centering
        \includegraphics[width=\linewidth]{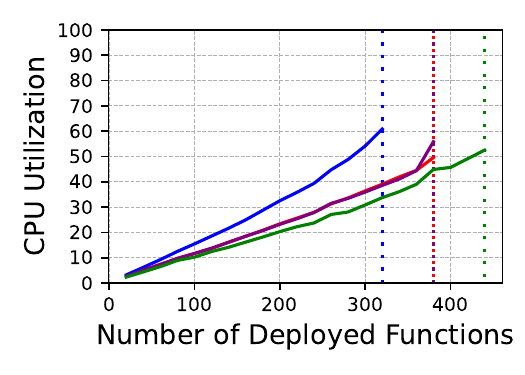}
        \caption{CPU Utilization.}
        \label{fig:eval-cpu-util}
    \end{subfigure}
    \begin{subfigure}[t]{0.49\linewidth}
        \centering
        \includegraphics[width=\linewidth]{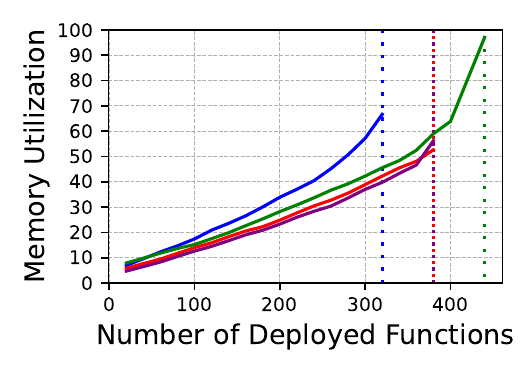}
        \caption{Memory Utilization.}
        \label{fig:eval-mem-util}
    \end{subfigure}\hfill
    \caption{End-to-End latency evaluation and resource utilization as deployment density scales. Each deployed function serves a trace sampled from the Azure Functions production dataset~\cite{shahrad:serverless}.
    % \todo{ETA: 10 worker 29th}
    }
    \label{fig:eval-resource-util}
    \vspace{-15pt}
\end{figure}

We begin with an end-to-end mixed-workload trace replay to show how \sys improves deployment density. 
We run a mix of functions, and each function can have multiple instances running concurrently in the cluster. We follow a synchronous autoscaling policy used by AWS Lambda~\cite{aws-lambda-scaling}, which adjusts the number of instances on demand for each function.
% \jy{, with 1 concurrent invocation per VM}. by DMI
Each VM is configured with 512MB of memory, and the compute budget is limited to 1 vCPU by Cgroup, based on the function configurations used in AWS Lambda~\cite{datadog}.
We measure slowdown (99th percentile latency normalized to the unloaded median latency) for each function as we sweep the number of deployed functions, until the geometric mean slowdown violates the SLO. Each function comes with a dedicated trace sampled from Azure Function traces that the load generator replays to its instances, which scale on demand.

Figure~\ref{fig:mix-trace-e2e} shows that Baseline sustains up to 320 deployed functions while meeting the target SLO, whereas \sys-TCP and \sys-Async sustain 380 and \sys sustains 440, respectively, corresponding to the deployment density gains of 18\% and 37\%, respectively. 

To explain these benefits, we analyze the cluster resource usage across the worker nodes.
Figures~\ref{fig:eval-cpu-util} and ~\ref{fig:eval-mem-util} show the averaged CPU and memory utilization as we sweep the load. To compare resource efficiency at a common operating point, we examine the largest scale Baseline can support: 180 functions. At that point, \sys-TCP reduces CPU and memory utilization by 35\% and 36\%, respectively, and \sys-Async reduces CPU and memory utilization by 36\% and 40\%, respectively, compared to Baseline, while \sys reduces CPU utilization by 44\% and memory utilization by 31\%.

Taken together, these results show that \sys serves more functions under the same latency target while using worker resources more efficiently. 
The gain comes from two complementary effects: First, \sys-TCP removes the duplicated communication fabric from each tenant VM and amortizes it in a shared backend
which uses the Go programming language to execute the cloud I/O SDK, which is more efficient in terms of CPU cycles than Python.
Second, \sys further reduces host CPU cycles by replacing TCP with RDMA. 
TCP operations constantly engage the host user and the host kernel, whereas RDMA bypasses the host kernel during communication and directly maps the payload to a shared memory region, resulting in fewer CPU cycles per transfer than TCP.
Also, \sys-Async shows lower memory utilization than \sys-TCP due to asynchronous output and early VM release, which increases VM utilization.

\subsection{Efficiency Analysis \& Ablation Study}
\label{sec:eval_ablation}

To explain the deployment density gains observed in the end-to-end study, we conduct an ablation study and an efficiency analysis, revisiting the defined density constraints: CPU and memory. We analyze how \sys's compute and I/O separation, as well as latency-overlapping optimizations, reduce warm-path CPU overhead (\S\ref{sec:eval_cpu_cycles}) and the memory footprint (\S\ref{sec:eval_mem}), and quantify the implications for cold-start latency (\S\ref{sec:eval_cold}).

\subsubsection{CPU Cycles}
\label{sec:eval_cpu_cycles}

\begin{figure}
    \centering
    \includegraphics[width=1.\linewidth]{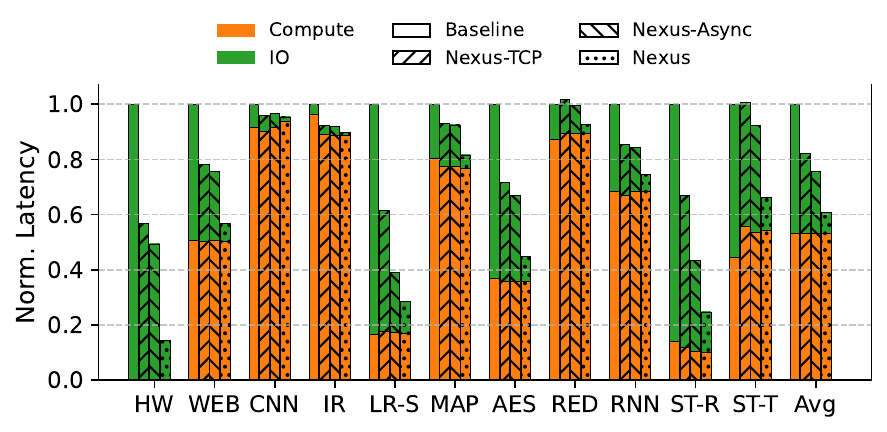}
    \caption{Warm latency across vSwarm workloads normalized to Baseline. \sys reduce guest-side I/O processing.}
    \label{fig:warm-latency}
    \vspace{-12pt}
\end{figure}

We first evaluate the impact of compute and I/O decoupling on warm execution latency using the same set of vSwarm functions as described in \S\ref{sec:method}.
We measure unloaded latency by deploying a single function instance and repeatedly sending a request, discarding the first, for each workload.
\cz{Why is this figure used in the CPU cycles?}
Figure~\ref{fig:warm-latency} shows that, compared to Baseline, \sys-TCP, \sys-Async, and \sys reduce warm latency by 19\%, 22\%, and 39\% on average, respectively. 
The benefit is strongly workload-dependent, favoring the I/O-intensive workloads, which benefit from the optimized I/O data path via the shared memory transport of \sys. I/O-heavy workloads, such as Linear Regression-Serving~(LR-S) and Stack Training-Reducer~(ST-R), improve the most, with latency reductions of 75\% and 78\%, respectively, whereas a compute-heavy workload, such as the CNN-based image recognition workload, improves by only 8\%.

\begin{figure}
    \centering
    \includegraphics[width=1.\linewidth]{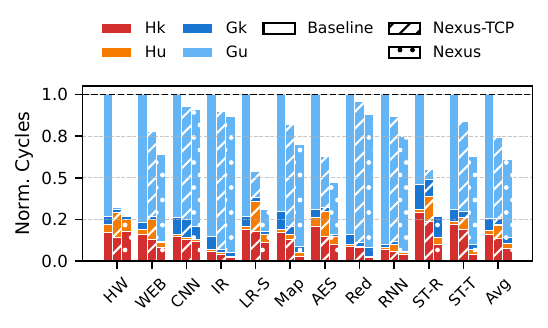}
    \caption{CPU cycles breakdown for each workload under the three studied systems, normalized per invocation: Baseline(left), \sys-TCP(center) \& \sys(right) across \texttt{Hk(host kernel)}, \texttt{Hu(host user)}, \texttt{Gk(guest kernel)} and \texttt{Gu(guest user)} spaces.
    }
    \label{fig:workload-cycles}
    \vspace{-12pt}
\end{figure}

\begin{figure}
    \centering
    \includegraphics[width=1.\linewidth]{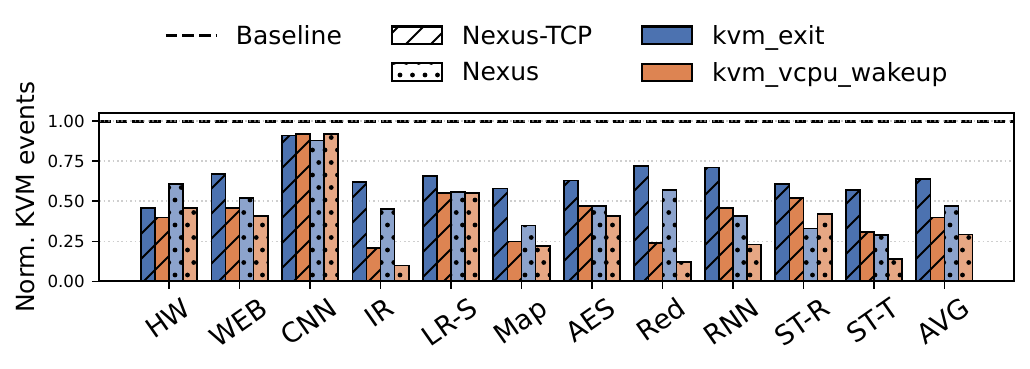}
    \caption{\texttt{kvm exit} and \texttt{kvm vcpu wakeup} event rates across vSwarm workloads normalized per invocation. \sys reduces both rates compared to baseline (\textbf{1.0}).}
    \label{fig:workload-kvm}
    \vspace{-12pt}
\end{figure}

To identify the source of these gains, we collect CPU cycle breakdowns and KVM activity measurements for each function under load. Here, to break down the cycle distribution across the user/kernel/guest/host layers, we run a separate experiment for each function, with several instances of the same function serving invocations. To minimize noise from the control plane and instance creation, we set the number of function instances to a fixed value.
To collect the CPU cycle breakdown per invocation, we use the $perf$~\cite{perf} tool to measure them across the entire node, using a $perf$ argument to break the collection into guest and host user and kernel space, and report them normalized to the baseline. For KVM activity, we use the $perf$-$kvm$ tool and deduct per invocation. The results are normalized to the baseline.

Figure~\ref{fig:workload-cycles} shows that \sys reduces total CPU cycles per request by 37\%, on average. This reduction is accompanied by a 28\% average drop in guest-user cycles. 
The largest savings again appear in the I/O-intensive workloads as presented before (LR-S, ST-R, and ST-T), which also exhibit the sharpest declines in KVM activity.
Figure~\ref{fig:workload-kvm} shows a 53\% drop in KVM exits and a 70\% drop in KVM vCPU wakeups, on average, which correlate well with the warm latency reductions in Figure~\ref{fig:warm-latency}.
\sys further cuts host-kernel cycles by 54\% relative to \sys-TCP because of RDMA bypassing the standard networking stack. 
Although host user-space cycles increase by 71\%, this increase reflects work moving out of the guest and into \sys's shared backend,  where it can be executed more efficiently because it's written in Go, so the total number of cycles still falls. 
However, compute-intensive workloads, e.g., CNN, benefit less from \sys, since they are highly dominated by computation during execution.

Overall, these results show that the warm-path latency improvement comes from eliminating redundant guest-side I/O, collapsing much of the guest-host virtual devices' communication path into a shared-memory communication path between VMs and \sys's backend, and further reducing kernel involvement when RDMA replaces TCP, since RDMA bypasses the traditional networking stack.

\subsubsection{Memory Footprint}
\label{sec:eval_mem}
\begin{figure}
    \centering
    \includegraphics[width=1.\linewidth]{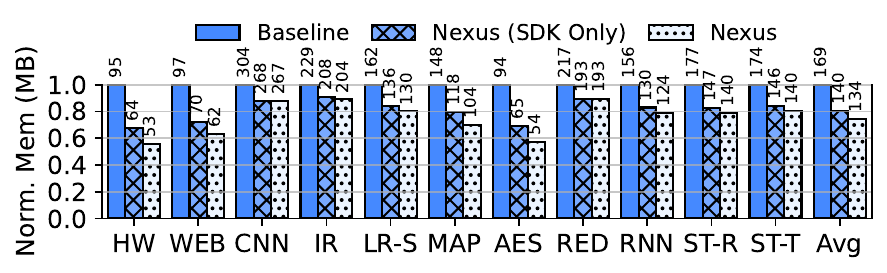}
    \caption{Per-function instance memory footprint across vSwarm workloads, normalized to Baseline. \sys reduces per-VM memory footprint by consolidating the communication fabric out of the VM to \sys's backend.
    }
    \label{fig:per-workload-memory}
    \vspace{-12pt}
\end{figure}

\begin{figure}
    \centering
    \includegraphics[width=1.\linewidth]{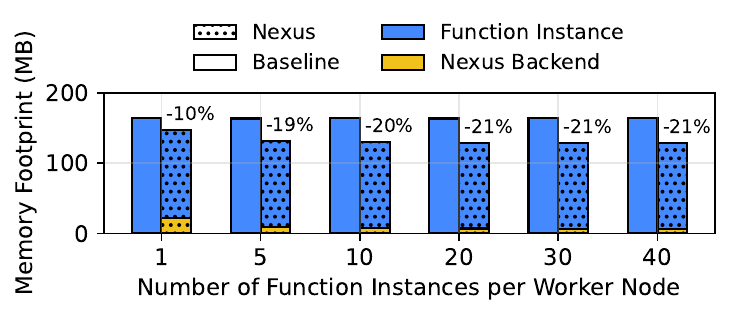}
    \caption{Worker node memory footprint breakdown across various deployment densities. \sys amortized the shared communication fabric on the \sys's backend, consistently reducing the footprint by 10-21\%.
    % \dmi{caption is not descriptive}
    % \dmi{shrink vertically by 2x}
    }
    \label{fig:mix-trace-memory}
    \vspace{-12pt}
\end{figure}

We next show that offloading the communication fabric raises the memory ceiling that limits deployment density, as well as CPU cycles. 
Figure~\ref{fig:mem-footprint} evaluates this at the instance level by separating the optimizations into two additive configurations: \sys (SDK Only) offloads the cloud SDK, but not RPC, to the \sys backend, whereas \sys offloads both cloud I/O SDK and platform RPC layer. We did not add \sys-Async to this experiment as it has the same memory footprint as \sys. 

Across all workloads, per-instance memory drops from 169 MB in Baseline to 140 MB with SDK-only offload and to 134 MB with communication-fabric offload, corresponding to average reductions of 17\% and 20\%, respectively. 
Even functions that rely heavily on large libraries, such as CNN/RNN and LR-S, which use PyTorch and Pandas, respectively, consistently shed about 30–40 MB. 
These savings arise because Baseline carries communication fabric within every VM, whereas \sys consolidates that state in the \sys backend shared among all the co-resident VMs, leaving only a thin frontend in the guest.

At the node level, the same trend persists as the number of co-resident instances grows. Figure~\ref{fig:mix-trace-memory} shows that total node memory remains about 21\% lower as we scale the number of function instances per worker. This consistency indicates that the backend cost is amortized across tenants rather than growing in proportion to the number of VMs. Importantly, the shared \sys backend, written in Go, is more memory-efficient than the Python library running inside the baseline VMs, so remoting services' SDK API to \sys is sensible if at least one instance per node uses that service.

\subsubsection{Cold Latency Breakdown}
\label{sec:eval_cold}

\begin{figure}
    \centering
    \includegraphics[width=1.\linewidth]{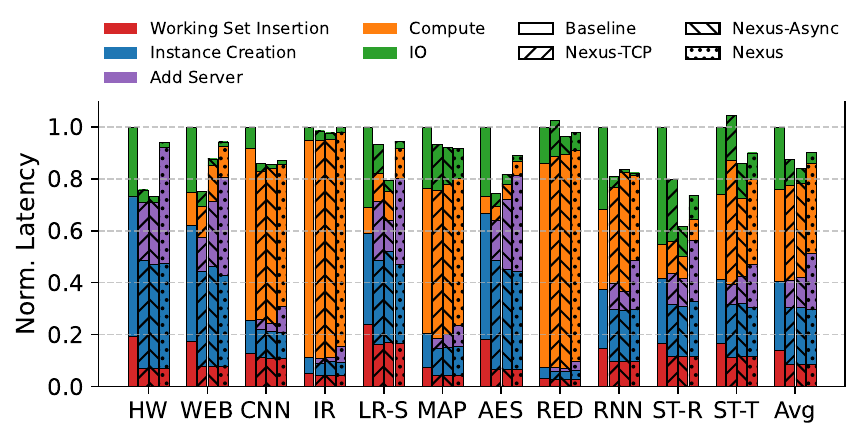}
    \caption{
    Normalized cold-start latency breakdown across the vSwarm suite. \sys reduces total cold-start delays by overlapping input prefetching with VM creation and shrinking the mandatory snapshot footprint.
    }
    \label{fig:cold-latency}
     \vspace{-15pt}
\end{figure}

\begin{figure}
    \centering
    \includegraphics[width=1\linewidth]{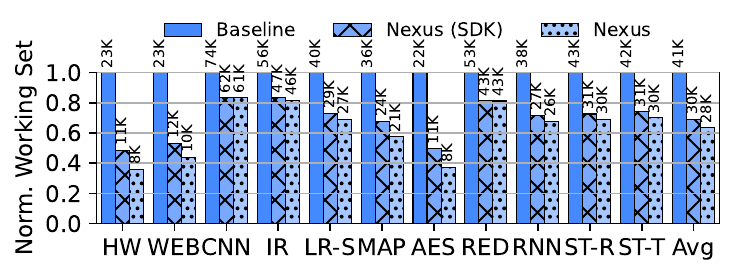}
    \caption{Snapshot working set size in pages during snapshot restoration~\cite{ustiugov:reap}. \sys drastically reduces the number of memory pages the hypervisor reads from disk.
    }
    \label{fig:workload-wss}
     \vspace{-12pt}
\end{figure}

We next analyze cold-start latency to understand how \sys reduces it by invoking functions one at a time. With instrumentation, we capture the latency breakdown (Figure~\ref{fig:cold-latency}) and the number of working set pages retrieved during the VM restoration from a snapshot (Figure~\ref{fig:workload-wss}).
Figure~\ref{fig:cold-latency} shows that \sys reduces cold-start latency by 10\% on average relative to Baseline, particularly in working set insertion time and I/O processing.

The first reason for the speedup is a 40\% reduction in working-set insertion time. Figure~\ref{fig:workload-wss} explains why: by offloading the communication fabric out of the VM, \sys reduces the working set of guest memory pages by 31\%, on average, allowing the hypervisor to fetch fewer pages during restoration, which accelerates it.

The second reason is the reduction in input retrieval and writeback time (I/O) on the critical path.
\sys-TCP reduces the I/O component by 58\% due to faster I/O processing, as for warm invocations~(\S\ref{sec:eval_cpu_cycles}). 
\sys-Async further reduces I/O processing time by 75\% by overlapping I/O with instance restoration and initialization, and moving writeback off the critical path (\S\ref{sec:design}), in contrast to the baseline, where VM creation, compute, and I/O processing are serialized.
Finally, \sys reduces I/O processing by 81\% by accelerating payload transfers with RDMA, bypassing the kernel. 

These gains are partially offset by \sys backend's establishing and managing connections on behalf of the VMs, 
reflected in \texttt{Add Server} category in Figure~\ref{fig:cold-latency}, which are subject to further optimizations. Specifically RDMA connection setup that contributes to increase this category the most.
Nevertheless, \sys still achieves a net 10\% reduction in cold-start latency, on average, by enabling faster, leaner restoration and breaking the baseline’s strict restore-then-fetch serialization.

\subsection{Comparison with a Lightweight WASM Hypervisor: Faasm Case Study}
\label{sec:eval_faasm}

\begin{figure}
    \centering
    \includegraphics[width=1.\linewidth]{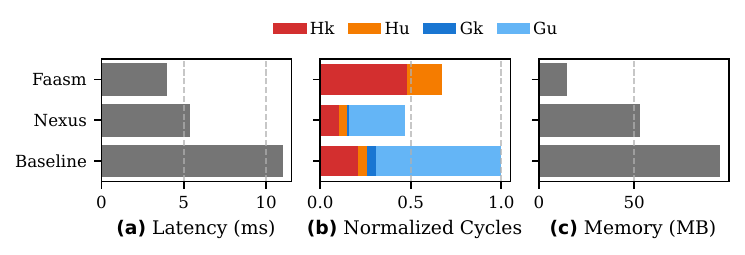}
    \caption{Execution time, per-invocation CPU cycles breakdown, and memory footprint of AES encryption workload under the 3 studied systems: Baseline, \sys, and Faasm.
    }
    \label{fig:eval-faasm}
     \vspace{-15pt}
\end{figure}

Finally, we compare \sys efficiency with Faasm~\cite{shillaker:faasm}, a state-of-the-art WASM-based hypervisor, quantifying the gap between \sys and its ecosystem-incompatible alternatives, such as Dandelion~\cite{kuchler:dandelion}, whose efficient runtime is also based on WASM, using the compute-I/O balanced AES function.\footnote{Faasm dropped support for Python and its module ecosystem due to the maintenance challenges they impose~\cite{faasm-python-supp}; hence, in Faasm, we instead use a C++ benchmark version, comparing it to the corresponding AES benchmark running in \sys. }
Comparing the latency and CPU cycles per invocation collected with \texttt{perf} under medium load, one can see that Faasm and \sys differ by a moderate 20-25\% (Figures \ref{fig:eval-faasm}a and \ref{fig:eval-faasm}b).\footnote{The high kernel cycle usage in Faasm is caused by the large amount of time spent in page faults triggered during Faasm's control plane execution (Faabric), which bootstraps WASM sandboxes (see the flamegraph in the \emph{Supplementary material}). This is also why Faasm's total cycles exceed \sys cycles despite the lower latency.}
However, since given \sys still boots a general-purpose VM with a guest OS, it still uses 3.5$\times$ more memory than Faasm (\ref{fig:eval-faasm}c), which alone may not justify the WASM porting and maintenance challenges~(\S\ref{sec:back_alternatives}).

%% file: sec/5_related.tex
\section{Related Work}

\noindent \textbf{Serverless Sandboxing and Lightweight Isolation.}
Production platforms rely on conventional VMs~\cite{agache:firecracker, randazzo:kata} for strong isolation, but duplicating the guest OS and communication fabric in every instance limits deployment density.
Unikernels and library OSes~\cite{kuenzer:unikraft,kivity:osv,shen:xcontainers,you:alloystack} shrink footprints by collapsing the execution environment, while single-address-space designs~\cite{li:jord,kotni:faastlane} eliminate inter-function isolation within workflows.
Other approaches abandon standard virtualization entirely via WebAssembly~\cite{shillaker:faasm, cloudflare_workers}, lightweight threads~\cite{dukic:photons}, or kernel-bypass execution~\cite{fried:junction,wanninger:virtines}.
All of these sacrifice compatibility with the FaaS programming model, high-level runtimes, or POSIX~\cite{wasm-limit,wasm-gap,wasm-state}.
Orthogonally, cold-start optimizations~\cite{ustiugov:reap, du:catalyzer, liu:fastiov, guo:vpri, margaritov:ptemagnet} speed up snapshot restoration and memory management but do not address snapshot bloat caused by the per-VM communication fabric.
\sys retains a full KVM-based VM and POSIX environment but extracts only the duplicated communication fabric, reducing both steady-state overhead and snapshot size while compounding with existing cold-start techniques.

\noindent \textbf{API Remoting and I/O Offloading.}
Splitting functionality across execution boundaries is well established, from datacenter disaggregation~\cite{shan:legoos} to accelerator remoting~\cite{yu:ava,strati:orion}.
In networking, LineFS~\cite{kim:linefs}, Junction~\cite{fried:junction}, and Palladium~\cite{qi:palladium} offload RPC, TCP, or file-system processing to host threads, SmartNICs, or DPUs---operating at the transport or storage layer and typically requiring specialized hardware.
\sys remotes at the cloud SDK API boundary instead, a higher-level, stable interface that lets it offload request construction, authentication, serialization, and connection management on commodity hardware, while remaining orthogonally compatible with hardware-accelerated transports.

\noindent \textbf{Serverless Data Management and I/O Separation.}
Several systems redesign serverless data paths and state management. Pocket~\cite{klimovic:pocket} provides ephemeral storage tiers, Cloudburst~\cite{sreekanti:cloudburst} co-locates caches with executors, OFC~\cite{mvondo:ofc} caches intermediate data, Boki~\cite{jia:boki} offers shared logs, and Nightcore~\cite{jia:nightcore} optimizes inter-function RPCs. These systems fundamentally optimize backend storage or data-passing abstractions, yet they natively retain the heavyweight communication fabric coupled within each isolated guest VM. \sys is entirely orthogonal and complementary to these approaches; it transparently offloads the transport layer of these optimized backends to achieve even higher efficiency.

Dandelion~\cite{kuchler:dandelion} also structurally separates compute from I/O, but requires developers to manually rewrite applications, forfeiting POSIX and mature ecosystem compatibility. In contrast, \sys achieves \emph{transparent} separation at the standard provider SDK boundary. By shifting the maintenance of interception stubs to the cloud provider, \sys extracts the I/O tax from the KVM sandbox without requiring any user code modifications, securing high efficiency while preserving legacy compatibility.

% \noindent \textbf{Serverless Data Management and I/O Separation.}
% Several systems redesign serverless data paths. Pocket~\cite{klimovic:pocket} provides ephemeral storage tiers, Cloudburst~\cite{sreekanti:cloudburst} co-locates caches with executors, OFC~\cite{mvondo:ofc} caches intermediate data, Boki~\cite{jia:boki} offers shared logs, and Nightcore~\cite{jia:nightcore} optimizes inter-function RPCs.Dandelion~\cite{kuchler:dandelion} goes furthest by structurally separating compute from I/O for independent elasticity, but requires manual application decomposition via a new API. These systems redesign backend storage or data-passing abstractions yet retain the coupled per-VM communication fabric. \sys is complementary. It transparently removes the per-VM fabric overhead without requiring a new programming model, and can benefit from any of these optimized backends.

%% file: sec/7_conclusion.tex
\section{Conclusion}

Serverless computing has long operated under an assumption: strict multi-tenant isolation requires packing the entire execution and infrastructure stack into every individual sandbox. Through \sys, we demonstrate that this tightly coupled architecture is a fundamental bottleneck to cloud efficiency. By cleanly separating application logic from I/O and offloading the latter to a shared host backend, \sys redefines the serverless virtualization boundary. We show that \sys increases deployment density by 37\%.